\documentclass[12pt,preprint]{aastex}

\shorttitle{EARLY AFTERGLOW OF GRB 050525A}
\shortauthors{Shao & Dai}

\begin{document}
\title{A Reverse-Shock Model for the Early Afterglow of GRB 050525A}

\author{L. Shao and Z. G. Dai}
\affil{Department of Astronomy, Nanjing University, Nanjing 210093, China;
dzg@nju.edu.cn}

\begin{abstract}
The prompt localization of gamma-ray burst (GRB) 050525A by {\em Swift} allowed
the rapid follow-up of the afterglow. The observations revealed that the
optical afterglow had a major rebrightening starting at $\sim 0.01$ days and
ending at $\sim 0.03$ days, which was followed by an initial power-law decay.
Here we show that this early emission feature can be interpreted as the reverse
shock emission superposed by the forward shock emission in an interstellar
medium environment. By fitting the observed data, we further constrain some
parameters of the standard fireball-shock model: the initial Lorentz factor of
the ejecta $\gamma_0>120$, the magnetic energy fraction
$\epsilon_B>4\times10^{-6}$, and the medium density $n<2\,{\rm cm^{-3}}$. These
limits are consistent with those from the other very-early optical afterglows
observed so far. In principle, a wind environment for GRB 050525A is
disfavored.
\end{abstract}

\section{Introduction}
The gamma-ray burst (GRB) 050525A is a bright brief flash of gamma-ray
radiation detected by the {\em Swift} Burst Alert Telescope (BAT) on 25 May
2005 00:02:53 (UT) \citep{b05}. It showed two peaks, with a duration of
$T_{90}=8.8\pm0.5 \,{\rm s}$ in the 15-350 keV band \citep{m05,c05b}. Fitting to
the Band spectral model yields a low-energy photon index of $\alpha=1.0\pm 0.1$
and a peak energy of $E_p=79\pm4$ keV \citep{c05b}. The fluence in the 15-350
keV band is $(2.0\pm0.1)\times10^{-5} \,{\rm ergs\,\,\rm cm^{-2}}$. This burst
was also detected by other on-board instruments such as INTEGRAL and
Konus-Wind, with peak fluxes of $\sim 3.2\times10^{-6}\,{\rm
ergs\,\,cm^{-2}\,\,s^{-1}}$ \citep{g05} and $\sim 8.7\times10^{-6}\,{\rm
ergs\,\,cm^{-2}\,\,s^{-1}}$ \citep{g05b}, respectively. The fluences were $\ge
1.2\times10^{-5}\,{\rm ergs\,\,cm^{-2}}$ in 20-200 keV (for the 12 second
integration time) by INTEGRAL and $\sim 7.8\times10^{-5}\,{\rm
ergs\,\,cm^{-2}}$ by Konus-Wind in 20-1000 keV (for the 11.5 second integration
time), respectively. The time-integrated spectrum shows a peak energy $E_p =
84.1 \pm 1.7 {\rm keV}$ by Konus-Wind \citep{g05b}. About six minutes later,
the ROTSE III telescope in Namibia was able to obtain images, and detected an
optical afterglow of magnitude 14.7 \citep{r05a}. A presumed host galaxy was
measured with a redshift of z=0.606, based on both [O III] 5007 and $H_\beta$
emission and Ca H$\&$K and Ca I 4228 absorption \citep{f05}. Assuming
cosmological parameters $\Omega_M=0.27$, $\Omega_\Lambda=0.73$, and
$H_0=71\,{\rm km}\,{\rm s}^{-1}\,{\rm Mpc}^{-1}$, the isotropic-equivalent
gamma-ray energy is about $1.2\times10^{53} \,{\rm ergs}$.

The follow-up observations of GRB 050525A revealed that the optical afterglow
decayed as $\propto t^{-1.3}$ until about 0.01 days after the burst.
Subsequently the afterglow showed a major rebrightening starting at $\sim 0.01$
days and ending at $\sim 0.03$ days, and then decayed as $\propto t^{-1.0}$.
This early-time emission feature is similar to the behavior of external reverse
shock emission predicted by the standard fireball-shock model
\citep{mr97,mr99,sp99a}, which was first confirmed by the detections of early
prompt optical emission from GRB 990123 \citep{a99, sp99b}. The reverse shock
model was also suggested to interpret early optical emission of GRB 021004
\citep{f03a,kz03a} and GRB 021211 \citep{f03b,l03,w03,pk04b}. Besides, the
prompt optical-infrared emission from GRB 041219A was recently suggested to be
the internal shock emission, since a strong correlation between $\gamma$-ray
and optical-infrared signals was detected \citep{v05,b05b}, which further
confirmed the standard internal-external-shock model of GRBs. Mirabal et al.
(2005) reported a break at $\sim 0.4$ days in the optical afterglow light curve
of GRB 050525A. In this paper, we show that the optical afterglow in
$\sim 0.4$ days after the {\em Swift} trigger can be well understood due to the
reverse shock emission superposed by the forward shock emission.

\section{Optical Emission from Forward and Reverse Shocks}
\label{sec:emission} The standard model for GRBs and their afterglows is the
fireball-shock model (for recent review articles, see M\'esz\'aros 2002; Zhang
\& M\'esz\'aros 2004; Piran 2005). In this model, the prompt emission of GRBs
is ascribed to internal shocks and the long-term afterglow to external shocks.
The prompt or very early optical emission is therefore explained as a
consequence of the reverse component of the external shocks \citep{mr97,sp99b}.

For shock-accelerated slow-cooling electrons with a power-law distribution, the
typical spectrum of synchrotron emission is described by three power laws
\citep{spn98}: (1) $F_\nu=(\nu/\nu_m)^{1/3}F_{\nu,max}$ for $\nu<\nu_m$; (2)
$F_\nu=(\nu/\nu_m)^{-(p-1)/2}F_{\nu,max}$ for $\nu_m<\nu<\nu_c$; (3)
$F_\nu=(\nu_c/\nu_m)^{-(p-1)/2}(\nu/\nu_c)^{-p/2}F_{\nu,max}$ for $\nu>\nu_c$,
where $\nu_m$, $\nu_c$ and $F_{\nu,max}$ are the typical synchrotron frequency,
cooling frequency and peak flux, respectively. The parameter $p$ is the index
of the electron energy distribution. Here we ignore the synchrotron
self-absorption because its corresponding frequency may be much less than the
optical band of interest.

Both a forward shock and a reverse shock emerge when an ultrarelativistic cold
GRB ejecta with initial Lorentz factor of $\gamma_0$ sweeps up a stationary
cold interstellar medium: the forward shock propagates into the interstellar
medium and the reverse shock propagates back into the ejecta \citep{k94, sp95}.
The emission of the forward shock is characterized by \citep{spn98}
\begin{eqnarray}
\nu_{m,f}&=&4.6\times10^{11}\left({1+z\over2}\right)^{1/2}
\left({E_{iso}\over10^{53}\,{\rm ergs}}\right)^{1/2}
\left({\epsilon_B\over10^{-3}}\right)^{1/2}
\left({\epsilon_e\over10^{-1}}\right)^2\left(g_m\over0.087\right)t_d^{-3/2}\,\, {\rm Hz},
\end{eqnarray}
\begin{eqnarray}
\nu_{c,f}&=&5.8\times10^{16}\left({1+z\over2}\right)^{-1/2}
\left({E_{iso}\over10^{53}\,{\rm ergs}}\right)^{-1/2}\left({\epsilon_B\over10^{-3}}\right)^{-3/2}
\left(g_c\over0.128\right)n^{-1}t_d^{-1/2}\,\, {\rm Hz},
\end{eqnarray}
\begin{eqnarray}
t_{m,f}&=&1.1\times10^{-2}\left({1+z\over2}\right)^{1/3}
\left({E_{iso}\over10^{53}\,{\rm ergs}}\right)^{1/3}
\left({\epsilon_B\over10^{-3}}\right)^{1/3}\left({\epsilon_e\over10^{-1}}\right)^{4/3}
\left(g_m\over0.087\right)^{2/3}\nonumber \\ & & \times
 \left({\nu_R\over10^{15}\,{\rm Hz}}\right)^{-2/3}\,\, {\rm days},
\end{eqnarray}
\begin{eqnarray}
F_{\nu,max,f}&=&25\left({1+z\over2}\right)\left({E_{iso}\over10^{53} \,{\rm
ergs}}\right)\left({\epsilon_B\over10^{-3}}\right)^{1/2}
\left({D_L\over10^{28}\,{\rm cm}}\right)^{-2}\left(g_{max}\over2.44\right)n^{1/2}\,\,{\rm mJy},
\end{eqnarray}
where, $t_{m,f}$ is the critical times when the break frequency, $\nu_{m,f}$
crosses the observed frequency $\nu_R$, $\epsilon_B$ is the fraction of the
shock energy goes into the magnetic field, $\epsilon_e$ is the fraction of the
shock energy goes into the electrons, $g_m=(p-0.67)(p-2)^2/(p-1)^2$,
$g_c=(p-0.46)e^{-1.16p}$, $g_{max}=p+0.14$ \citep{gs02}, $E_{iso}$ is the
isotropic-equivalent kinetic energy, $n$ is the density of interstellar medium
in units of 1 ${\rm cm^{-3}}$, $z$ is the redshift of the burst, $D_L$ is the
corresponding luminosity distance, and $t_d$ is the observer's time in units of
1 day. The above equations are valid for $p>2$. Afterglows with the hard
electron spectrum of $1<p<2$ have been discussed by Dai \& Cheng (2001).

For reverse shocks, it is possible to get a simple analytic solution in two
limiting cases: thin shell and thick shell (which are corresponding to
Newtonian and relativistic reverse shock, respectively; Sari $\&$ Piran, 1995).
Since the reverse-shocked gas is separated with the forward-shocked gas by a
contact discontinuity, which keeps the equality of pressures and velocities in
both shocks, we can find the emission properties of reverse shock in the aid of
the correlations between forward and reverse shocks at the crossing time
$t_\times$ \citep{k00,kz03a}. First, since the reverse- and forward-shocked
gases have the same magnetic field and Lorentz factor at $t_\times$, the
cooling frequency of the reverse-shocked electrons is equal to that of the
forward-shocked electrons, $\nu_{c,r}(t_\times)=\nu_{c,f}(t_\times)$. Second,
the electron's random Lorentz factor of the reverse shock is smaller than that
of the forward shock by a factor of $\gamma_0/\gamma_\times^2$ (where
$\gamma_\times$ is the Lorentz factor of both shocked regions at $t_\times$),
the typical synchrotron frequency of the reverse shock is by a factor of
$\gamma_0^2/\gamma_\times^4$ less than that of the forward shock,
$\nu_{m,r}(t_\times)=(\gamma_0^2/\gamma_\times^4)\nu_{m,f}(t_\times)$. This is
basically valid for the thick shell case. As for a thin shell, it is more
realistic with the expression,
$\nu_{m,r}(t_\times)=(\gamma_{34,\times}-1)^2/(\gamma_\times-1)^2\nu_{m,f}(t_\times)$,
where
$\gamma_{34,\times}\simeq(\gamma_\times/\gamma_0+\gamma_0/\gamma_\times)/2$ is
the Lorentz factor of the shocked ejecta in the rest frame comoving with the
unshocked ejecta at the crossing time and $\gamma_\times\simeq\gamma_0/2$
\citep{zkm03,fw05}. Third, the peak flux $F_{\nu, max}$ of an emission region
is proportional to the electron number, the magnetic field, and the Lorentz
boost. Since at $t_\times$, the electron number of the ejected shell is larger
than that of the swept-up medium by a factor of $\gamma_\times^2/\gamma_0$, we
obtain $F_{\nu, max,r}(t_\times) =(\gamma_\times^2/\gamma_0)F_{\nu,
max,f}(t_\times)$. Therefore, we can directly write down the corresponding
characteristic variables of the reverse shock emission, at $t=t_\times$,
\begin{eqnarray}
\nu_{m,r}(t_\times)&=&4.6\times10^{11}\left({1+z\over2}\right)^{-1}\left({\epsilon_e\over10^{-1}}\right)^2
\left({\epsilon_B\over10^{-3}}\right)^{1/2}n^{1/2}\left(g_m\over0.087\right)
\nonumber \\ & & \times \left({\gamma_0\over10^{2.5}}\right)^2\,\,{\rm Hz},
\end{eqnarray}
\begin{eqnarray}
\nu_{c,r}(t_\times)&=&2.8\times10^{18}\left({1+z\over2}\right)^{-1}\left({E_{iso}\over10^{53}\,{\rm
ergs}}\right)^{-2/3}\left({\epsilon_B\over10^{-3}}\right)^{-3/2}n^{-5/6}
\nonumber \\ & & \times \left(g_c\over0.128\right)
\left({\gamma_0\over10^{2.5}}\right)^{4/3}\,\,{\rm Hz},
\end{eqnarray}
\begin{eqnarray}
F_{\nu,max,r}(t_\times)&=&10\left({1+z\over2}\right)\left({\epsilon_B\over10^{-3}}\right)^{1/2}n^{1/2}
\left({E_{iso}\over10^{53}\,{\rm ergs}}\right)\left({D_L\over10^{28}\,{\rm
cm}}\right)^{-2} \nonumber \\ & & \times
\left(g_{max}\over2.44\right)\left({\gamma_0\over10^{2.5}}\right)\,\,{\rm Jy},
\end{eqnarray}
for the thin shell (i.e. Newtonian) case, and
\begin{eqnarray}
\nu_{m,r}(T)&=&5.1\times10^{11}\left({1+z\over2}\right)^{-1}\left({\epsilon_e\over10^{-1}}\right)^2
\left({\epsilon_B\over10^{-3}}\right)^{1/2}n^{1/2}\left(g_m\over0.087\right)\left({\gamma_0\over10^{2.5}}\right)^2
\,\,{\rm Hz},
\end{eqnarray}
\begin{eqnarray}
\nu_{c,r}(T)&=&1.7\times10^{18}\left({1+z\over2}\right)^{-1/2}\left({\epsilon_B\over10^{-3}}\right)^{-3/2}
n^{-1}\left({E_{iso}\over10^{53}\,{\rm
ergs}}\right)^{-1/2}\left(g_c\over0.128\right) \nonumber \\ & & \times
\left({T\over10^2\,{\rm s}}\right)^{-1/2}\,\,{\rm Hz},
\end{eqnarray}
\begin{eqnarray}
F_{\nu,max,r}(T)&=&2.3\left({1+z\over2}\right)^{7/4}\left({E_{iso}\over10^{53}\,{\rm
ergs}}\right)^{5/4} \left({\epsilon_B\over10^{-3}}\right)^{1/2}n^{1/4}
\left(g_{max}\over2.44\right)\left({D_L\over10^{28}\,{\rm cm}}\right)^{-2}
\nonumber \\ & & \times \left({T\over10^2\,{\rm
s}}\right)^{-3/4}\left({\gamma_0\over10^{2.5}}\right)^{-1}\,\,{\rm Jy},
\end{eqnarray}
for the thick shell (i.e. relativistic) case. Here, the GRB duration T is given
by the shell width $\Delta_0/c$, according to the internal shock model. As
derived by Sari \& Piran (1999a) and Kobayashi (2000), the typical synchrotron
frequency and peak flux evolve as $\nu_{m,r}\propto t^6$, $F_{\nu,max,r}\propto
t^{3/2}$ at $t<t_\times$, and $\nu_{m,r}\propto t^{-54/35}$,
$F_{\nu,max,r}\propto t^{-34/35}$ at $t>t_\times$, for the thin-shell case;
$\nu_{m,r}=$constant, $F_{\nu,max,r}\propto t^{1/2}$ at $t<T$, and
$\nu_{m,r}\propto t^{-73/48}$, $F_{\nu,max,r}\propto t^{-47/48}$ at $t>T$, for
the thick-shell case. Before the crossing time, the cooling frequency evolve as
$\nu_{c,r}\propto t^{-2}$ for the thin-shell case and $\nu_{c,r}\propto t^{-1}$
for the thick-shell case; after the crossing time, $\nu_{c,r}$ turns into the
cutoff frequency due to synchrotron cooling of electrons without acceleration,
and it evolves the same as $\nu_{m,r}$ for both cases. According to the
emission properties shown above, we derive light curves of the emission from
the forward- and reverse-shocked regions and thus can constrain the model
parameters by fitting the early afterglow of GRB 050525A.

\section{Constraints on Parameters of GRB 050525A}
From the emission features described by equations (1), (2), (5), (6), (8), and
(9), we see that the optical band is commonly between the typical frequency
$\nu_m$ and cooling frequency $\nu_c$ in both forward and reverse shocks around
the crossing time. Thus, the optical light curve of GRB 050525A can be fitted
by the superposition of two broken power-laws: (1) forward shock:
$F_{\nu,f}\propto t_d^{1/2}$ for $t_d < t_{m,f}$, and $F_{\nu,f}\propto
t_d^{3(1-p)/4}$ for $ t_d > t_{m,f}$; (2) reverse shock: $F_{\nu,r}\propto
t_d^{3(2p-1)/2}$ for $t_d < t_\times$, and $F_{\nu,r}\propto t_d^{(-27p+7)/35}$
for $ t_d > t_\times$. These scaling laws are valid for the thin-shell case, in
which the reverse shock decelerates the shell insignificantly \citep{k00}.

\begin{figure}
\epsscale{.80} \plotone{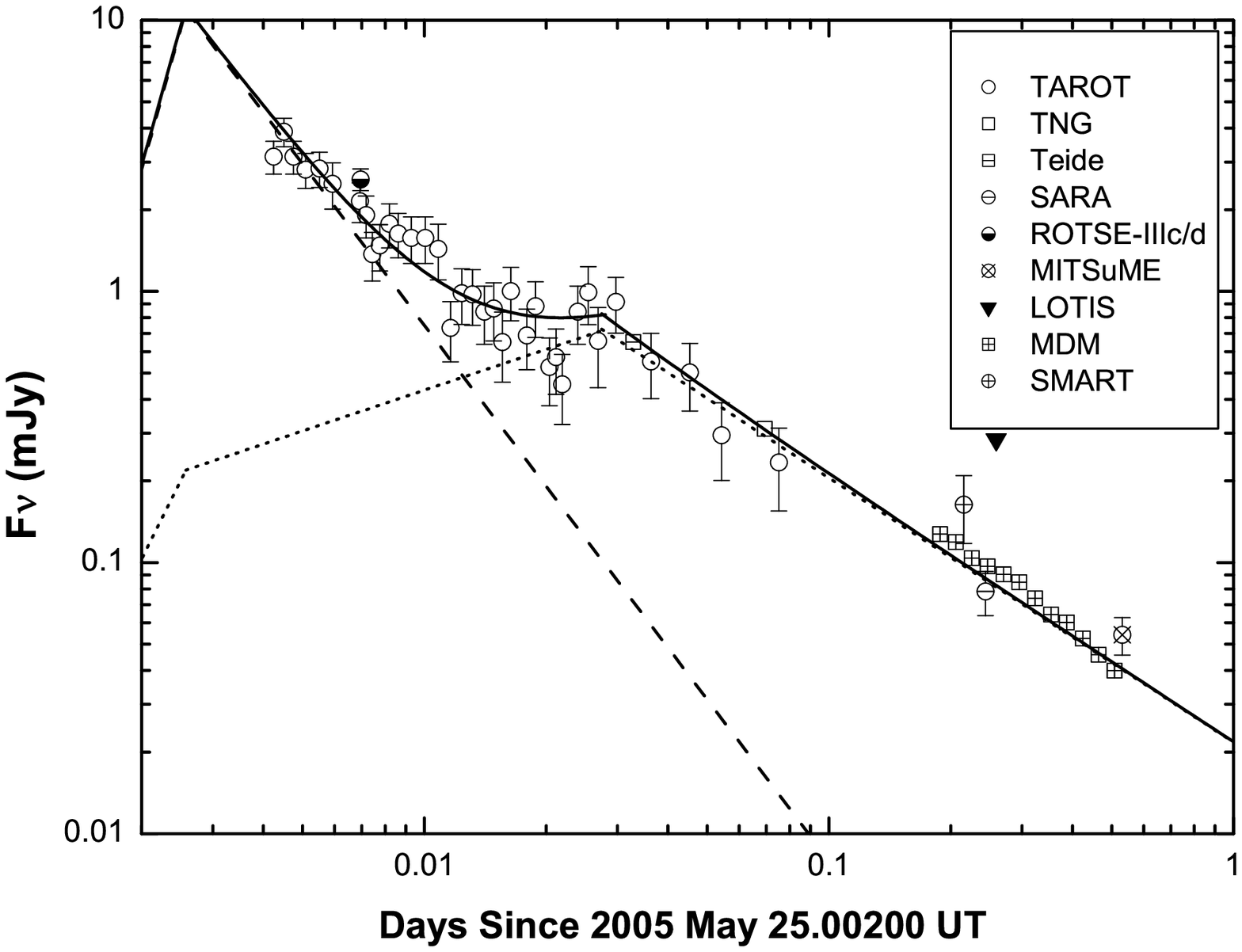} \caption{R-band afterglow light curve of
GRB050525A. The fitting curve (solid line) is generated by superposition of
forward shock emission (dotted line) and reverse shock emission (dashed line)
with $\epsilon_B=0.0025$, $\epsilon_e=0.2$, $n=0.003\,\,{\rm
cm^{-3}}$, $E_{iso}=10^{53}\,\,{\rm ergs}$, $p=2.3$, $\gamma_0=200$. These values are
basically consistent with the thin-shell case. Data are taken from the GRB
Coordinates Network (GCN):
\citet{m05b,k05a,tb05,h05,r05b,ytk05,mwp05,mbs05,cb05,k05b}.\label{fig1}}
\end{figure}

In Figure \ref{fig1} we present the R-band afterglow light curve of GRB 050525A
between May 25.0062 UT ($\Delta t\approx 0.004\,\,{\rm days}$) and May 25.5333
UT ($\Delta t\approx 0.531\,\,{\rm days}$). We find that, $t_{m,f}\approx
0.027\,\,{\rm days}$, $F_{\nu,max,f}\approx 0.719\,\,{\rm mJy}$, and $p\approx
2.3$ can fit the light curve well for the component of forward shock emission.
Given redshift $z=0.606$ \citep{f05}, the corresponding luminosity distance is
$D_L=1.1\times10^{28}\,{\rm cm}$. Using equations (3) and (4), we obtain
\begin{equation}
\epsilon_e\simeq8.3 n^{1/4}\left({E_{iso}\over10^{53}\,{\rm ergs}}\right)^{1/4}\,,
\end{equation}
and
\begin{equation}
\epsilon_B\simeq1.0\times10^{-5}n^{-1}\left({E_{iso}\over10^{53}\,{\rm ergs}}\right)^{-2}.
\end{equation}
Furthermore, from the flux inferred by the light curve of the reverse-shock
emission (depicted as a dashed line in Figure \ref{fig1}, where we see a flux
$F_{\nu,r}=0.75\, {\rm mJy}$ at $t=0.01\, {\rm days}$), using equations (5) and
(7), we further obtain
\begin{equation}
\left({E_{iso}\over10^{53}\,{\rm ergs}}\right)^{1.658}\left({\epsilon_B\over10^{-3}}\right)^{0.825}
\left({\epsilon_e\over10^{-1}}\right)^{1.3}n^{0.167}\left({\gamma_0\over10^{2.5}}\right)^{-2.965}
\simeq 5.9
\end{equation}
By simple algebra from equations (11), (12) and (13), we obtain a constraint on the initial Lorentz
factor of GRB ejecta,
\begin{equation}
\gamma_0=121\left({E_{iso}\over10^{53}\,{\rm ergs}}\right)^{0.112}\left({n\over1\,{\rm cm^{-3}}}\right)^{-0.112},
\end{equation}
which is weakly dependent on the isotropic-equivalent kinetic energy $E_{iso}$
and interstellar medium density $n$. Assuming the efficiency of energy
conversion is $\eta=0.5$, $E_{iso}$ is approximately equal to the detected
isotropic-equivalent $\gamma$-ray energy, i.e. $1.2\times10^{53}\,{\rm ergs}$.
Thus, we can obtain a mutual constraint between $\epsilon_B$ and $\epsilon_e$
from equations (11) and (12). By eliminating the medium density $n$, we have
$\epsilon_B=4\times10^{-6}\epsilon_e^{-4}$. Moreover, as a natural consequence
of $\epsilon_e<1$, we have $\gamma_0>116$, $\epsilon_B>4\times10^{-6}$, and
$n<2\,{\rm cm^{-3}}$. These constraints are more clear than the cases of GRB
990123 \citep{sp99b}, GRB 021004 \citep{kz03a} and GRB 021211 \citep{w03}. Our
preliminary fitting to the afterglow observations of GRB 050525A update to May
25.5333 UT ($\Delta t\approx 0.531\,\,{\rm days}$) is shown in Figure
\ref{fig1}.

Additionally, we also give a constraint on the parameters for the thick-shell
case. We can find from a fitting similar to that shown above that, the crossing
time of the reverse shock, i.e., the GRB duration $T$ is approximated by
\begin{equation}
T\simeq553\left({E_{iso}\over10^{53}\,{\rm ergs}}\right)^{0.06}
\left({n\over1\,{\rm cm^{-3}}}\right)^{-0.06}\left({\gamma_0\over10^{2.5}}\right)^{0.25}\,\,{\rm s},
\end{equation}
which is inconsistent with the observed $T_{90}\approx10\,\,{\rm s}$ \citep{m05,g05}.
This implies that GRB 050525A seems to be a thin-shell case.

\section{Discussion and Conclusions}
The optical afterglow of GRB 050525A rebrightened starting at $\sim 0.01$ days
and ending at $\sim 0.03$ days, which was followed by an initial power-law
decay. In this paper, we have shown that this early emission feature can
be interpreted as the reverse shock emission superposed by the forward shock
emission in an interstellar medium environment.
A good fitting to the observed light curve is shown in Figure 1. We
further find the initial Lorentz factor of the ejecta $\gamma_0>120$. Some
other model parameters are also constrained: the magnetic energy fraction
$\epsilon_B>4\times10^{-6}$ and the medium density $n<2\,{\rm cm^{-3}}$.
These limits are consistent with those from the other very-early optical
afterglows observed so far \citep{sp99b,kz03a,w03,pk04b,np05} and with the
values of the forward-shock parameters \citep{pk01}.

Our model is simplified under some assumptions: First, the reverse- and
forward-shocked regions have the same electron and magnetic energy fractions,
requiring that the ejecta is not initially magnetized. This requirement seems
to be relaxed for GRB 990123 \citep{f02,zkm03}. The interaction of a magnetized
 ejecta with its surrounding medium has been investigated in details \citep{fww04,zk05}.
Second, we neglect the effect of inverse Compton scattering (ICS) on the
reverse shock emission in \S 2. As shown by Wang, Dai, \& Lu (2001a, b), this
effect not only decreases the cooling frequency but also produces higher-energy
emission. However, this decrease does not affect the limits derived above
because the cooling frequency is much larger than the optical band at the times
of our interest. Third, the surrounding environment is assumed to be a uniform
interstellar medium, in which case two types of light curves are expected for
the reverse-forward-shock emission combination: a rebrightening light curve
(type I) and a flattening light curve (type II) \citep{zkm03}. Early afterglows
in wind environments have been studied by Chevalier, \& Li (2000), Wu et al.
(2003), Kobayashi \& Zhang (2003b), Panaitescu \& Kumar (2004b), and Zou et al.
(2005). Generally, a flattening light curve in optical band is expected in the
wind model \citep{kz03b,zwd05}, i.e., there is only type II light curve for the
reverse-forward-shock emission combination. Since the early optical afterglow
of GRB 050525A shows a rebrightening (type I) light curve, the reverse-forward
shock model in a wind environment for the early afterglow of GRB 050525A is not
favored.

The observed break at $t_j\simeq 0.4$ days in the optical afterglow light curve
of GRB 050525A \citep{mbs05} may be due to an ultrarelativistic jet. This
conclusion seems to be strengthened by Lulin Observatory around 17.7 hours
after the burst (at which the R-band magnitude is about $21.3\pm 0.1$, Chiang
et al. 2005). If so, the jet's half opening angle $\theta=0.051(t_j/0.4\,{\rm
d})^{3/8}(E_{iso}/10^{53}{\rm ergs})^{-1/8}(n/0.1\,{\rm
cm}^{-3})^{1/8}(\eta_\gamma/0.5)^{1/8}$ (Sari, Piran, \& Halpern 1999) and thus
the beaming-corrected gamma-ray energy $E_{jet}=1.3\times 10^{50}(t_j/0.4\,{\rm
d})^{3/4}(E_{iso}/10^{53}{\rm ergs})^{3/4}(n/0.1\,{\rm
cm}^{-3})^{1/4}(\eta_\gamma/0.5)^{1/4}$ ergs, which is by a factor of a few
smaller than the mean energy release found by Frail et al. (2001) for $n\sim
0.1\,{\rm cm}^{-3}$ and $\eta_\gamma\sim 0.5$. Even so, this energy still
satisfies the relation of Ghirlanda et al. (2004), $(E_{jet}/10^{50}{\rm
ergs})=(1.12\pm 0.12)[(1+z)E_p/100\,{\rm keV}]^{1.50\pm 0.08}$ (see Dai, Liang,
\& Xu 2004).

Finally, what we want to point out is that besides the reverse-shock model
discussed in this paper, the other plausible explanations of a rebrightening
light curve include the density-jump medium (Dai \& Lu 2002; Lazzati et al.
2002), pure Poynting-flux injection (Dai \& Lu 1998), baryon-dominated
injection (Rees \& M\'esz\'aros 1998; Granot, Nakar \& Piran 2003; Nakar, Piran
\& Granot 2003; Bj\"ornsson et al. 2004), two-component jet (Berger et al.
2003; Huang et al. 2003), neutron-fed fireball (Beloborodov 2003), temporal
fluctuation of $\epsilon_e$ and/or $\epsilon_B$ and departure of the electron
distribution from a power law (Panaitescu \& Kumar 2004). Multiwavelength
observations in the {\em Swift} era are expected to distinguish among these
possibilities.

\acknowledgments  We thank Y. Z. Fan and L. J. Gou for constructive comments, X.
F. Wu and Y. C. Zou for helpful discussions, and Y. F. Huang and X. Y. Wang for
a careful reading of our manuscript. This work was supported by the National
Natural Science Foundation of China (grants 10233010 and 10221001) and the
Ministry of Science and Technology of China (NKBRSF G19990754).

\end{document}